\documentclass[aps,prl,showpacs,twocolumn,longbibliography]{revtex4-2}

\usepackage{graphicx}
\usepackage[utf8]{inputenc}
\usepackage{dcolumn}
\usepackage{bm}
\usepackage{color}
\usepackage{amsmath} 
\usepackage{physics}
\usepackage{braket}
\usepackage{amssymb}
\usepackage{amsthm}
\usepackage{mathtools} 
\usepackage{amsfonts}
\usepackage{comment}
\usepackage{xcolor}
\usepackage{algorithm}
\usepackage{algorithmic}
\usepackage[caption=false]{subfig}

\begin{document}

\author{Baptiste Chevalier$^{1}$}
\email{baptiste.chevalier@keio.jp}

\author{Wojciech Roga$^{1}$}
 \email{wojciech.roga@keio.jp}
 
\author{Masahiro Takeoka$^{1,2}$}
 \email{takeoka@elec.keio.ac.jp}

\affiliation{%
$^{1}$Department of Electronics and Electrical Engineering, Keio University, 3-14-1 Hiyoshi, Kohoku-ku, Yokohama 223-8522, Japan} 

\affiliation{%
$^{2}$Advanced ICT Research Institute, National Institute of Information and Communications Technology (NICT), Koganei, Tokyo 184-8795, Japan
}

\date{\today}

\title{Compressed sensing enhanced by quantum approximate optimization algorithm}

\begin{abstract}
We present a framework to deal with a range of large scale compressive sensing problems using a quantum subroutine. We apply a quantum approximate optimization algorithm (QAOA) to support detection in a sparse signal reconstruction algorithm: matching pursuit. The constrained optimization required in this algorithm is difficult to handle when the size of the problem is large and constraints are given by unstructured patterns. Our framework utilizes specially designed structured constraints that are easy to manipulate and reduce the optimization problem to the solution of an Ising model which can be found using Ising solvers. In this research, we test the performance of QAOA for this purpose on a simulator of quantum computer. We observe that our method can outperform reference classical methods. Our results explore a promising path of applying quantum computers in the compressive sensing field.
\end{abstract}

\maketitle

\section{Background and Motivations} 

Compressive sensing has been intensely studied in recent years since the seminal papers \cite{Candes2,Donoho,Candes1,Baraniuk,Candes3} with significant impact on many practical technology fields such as, for instance: natural and medical imaging, compressed speech, audio and video, radar, communication, spectroscopy, matrix completion, big data, and quantum state tomography \cite{romberg, mihajlovic, lustig, gross}. In general terms, it is a procedure that allows one to find the original signal based on a number of linear measurements (filters) which is much smaller than the size of the signal. The procedure works when the signal is known to be sparse in some basis called the sparsity basis. If the number of measurements is appropriate, of the order of the number of non-zero elements in the sparse representation of the original signal, this signal appears as the vector in the sparsity basis which is the most sparse among those that fit the constraints. In standard approaches, this vector is found by a convex optimization procedure \cite{mallat,davis,Candes2,gradientpursuit,blumensath2,Draganic,zhang} with constraints given by the measured values and the measurement filter patterns. The convex optimization is an efficient procedure in terms of the speed of approaching the solution, however, in each step, it requires algebraic manipulations of the vectors of patterns which can be prohibitive if the size of the problem becomes comparable to big data \cite{Cevher}. Indeed, standard compressive sensing approaches use random vectors as the measurement pattern vectors. For them, if their size is too large, even the storage in the computer memory becomes challenging.  

At the same time we observe emerging quantum technology, including fault tolerant, noisy intermediate-scale, universal and dedicated quantum computers which allow for new approaches in dealing with large scale signals. In recent years we witness strong progress in developing both hardware and software \cite{Steane, Ladd, Gyongyosi, Huang, Albash, Flamini, Bruzewicz}. Moreover, a lot of effort has been spent showing quantum advantages over classical algorithms in various fields \cite{Arute, Wu, Zhong, Wang, Harrow, Bravyi,Havlicek,Cao, Hangleiter}. In the present research, we are interested in advantages that actual or potential quantum computers can offer to enhance compressive sensing especially in the regime of large scale problems.
Quantum computer applied to compressive sensing should be able to deal with large scale problems bringing improvement in both the running time and the quality of the solution with the ability to deal with many measurements patterns. Also showing an advantage of quantum computers in the field of compressive sensing opens the way to a whole new branch of applications for the latter.

The idea of applying quantum Ising solvers to greedy algorithms and compressive sensing, however so far not intensely explored, appeared already in the literature \cite{Ayanzadeh, Ayanzadeh2, roga_classical_2020,jacob_franck-condon_2020}. In particular, in \cite{Ayanzadeh} the authors were looking for the configuration of a number of spins with a few excitations which correspond to solutions of a posed Ising problem with sparse spectrum. In the regime of large signals, this approach would be infeasible due to the amount of resources required. 
The challenges faced by compressive sensing in the regime of large scale sparse signals have been already noticed in  \cite{roga_classical_2020,jacob_franck-condon_2020}. In those works, appropriately designed structured patterns of measurements have been proposed instead of the random patters of standard compressive sensing procedures. That removed the problem of memory overload. High complexity algebraic manipulations were also eliminated when the authors focused on first order iterative algorithms known as matching pursuit. The bottleneck of this method consists in finding the vector from a given library the most similar to a current residue vector. For the proposed patterns of measurements the library consisted of still large set of vectors, however the most similar vector to the current residue coincided with the solution of an Ising Hamiltonian. The authors considered only relatively small number of measurement patterns that made the corresponding Ising Hamiltonian the nearest neighbor one. Therefore, it could be efficiently solved by classical methods. 

However this approach worked only when the number of non-zero entries in the original sparse vector was small strongly limiting the proposed scheme. Including problems of smaller sparsity would require more constraints making the corresponding Ising Hamiltonian more general including non-local and many sites interaction. In general, these Hamiltonians can not be solved efficiently by polynomial complexity class algorithms leaving however space for solvers that can venture the field of problems that in the worst case are NP-complete. In this research we are interested in this extension of the method of \cite{roga_classical_2020,jacob_franck-condon_2020}.  

In this paper we present a general method to use quantum algorithms as a subroutine of compressed sensing algorithm and in this way apply it to a larger range of large scale sparse problems that can be tractable. We start with an introduction to compressive sensing and near term intermediate-scale quantum (NISQ) devices \cite{Preskill, Bharti, Benedetti, GarciaPerez}. In the second part, we present the framework that we use to deal with compressive sensing scenario. Finally, we focus on testing QAOA as an Ising solver in our approach and discuss results and comparison with alternative methods.

\section{Introduction}

Among dedicated quantum computers quantum annealers \cite{kadowaki_quantum_1998,Venturelli,Boixo,farhi_quantum_2000} realizing the so-called Adiabatic Quantum Computation \textbf{AQC} \cite{kadowaki_quantum_1998,Albash} are in the relatively advanced stage of technology development. Typical problems they solve consist of optimization of certain functions of many binary variables. A specific task needs to be mapped onto a physical system of quantum spins in such a way that the solution is the ground state of a Hamiltonian $H_p$ of the system that is called the problem Hamiltonian. In AQC, one also needs Hamiltonian $H_0$ 
the ground state of which is known and easy to prepare. Then, \textit{quantum adiabatic theorem} \cite{kadowaki_quantum_1998} guarantees evolution from the ground state of $H_0$ to the ground state of $H_p$ -- the solution of the problem -- if the evolution is performed slowly enough.
The efficiency of Adiabatic Quantum Computation has been studied, for instance, by Feynman \cite{feynman_quantum_1986} before being proven as efficient as the equivalent quantum computation in gate model \cite{aharonov_adiabatic_2005}. However the discussion on the quantum advantage on nowadays quantum annealers in particular cases is still ongoing \cite{Albash2,Heim}.

According to the adiabatic theorem, two Hamiltonians $H_0$ and $H_P$ should not commute (should be diagonal in different bases). 
The problem Hamiltonian $H_P$ is often chosen as an Ising Hamiltonian expressed by Pauli spin operators $\sigma^Z$ since it is easy to map any diagonal matrix, and hence any 
function, into it. Hamiltonian $H_0$ is usually taken as the sum of local Pauli operators $\sigma^X$, i.e., $H_0 = \sum_i \sigma^X_i$.

\begin{figure}[h]
    \centering
    \includegraphics[scale=0.25]{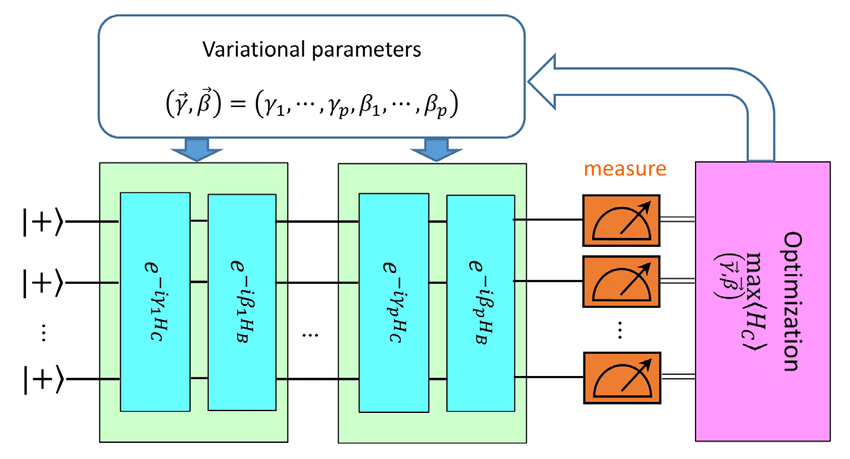}
    \caption{Schematic diagram of Quantum Approximation Optimization Algorithm.}
    \label{fig:QAOA}
\end{figure}

Quantum Approximation Optimization Algorithm \textbf{QAOA} proposed in 2014 \cite{farhi_quantum_2014, boulebnane_solving_2022} is a type of Variational Quantum Algorithm (VQA) using a specific Ansatz inspired by AQC. In VQA \cite{cerezo_variational_2021}, a parameterized multi-parameter circuit (also called Ansatz) is used. The circuit generates a state $\ket{\psi(\vec \gamma, \vec \beta)}$ in which the expectation value $E(\vec \gamma, \vec \beta)$ of a Hamiltonian is computed (classically or directly on a quantum computer) and used to evaluate an objective function. Using a classical optimization algorithm, one can find the best parameters for the circuit to produce the ground state of the Hamiltonian. 
In QAOA, we have an optimization problem which can be seen as a function $h(z)$ to be minimized (resp. maximized). The problem Hamiltonian associated with this function can be written as
\begin{equation}
    H_p = \sum_z h(z) \ket{z}\bra{z},
\end{equation}
which is diagonal in some basis $\{\ket{z}\}$. Similarly to the AQC, in the QAOA Ansatz we use two non-commuting Hamiltonians: the problem, $H_p$, and the mixing Hamiltonian $H_0$. The quantum circuit consists in applying unitary evolution driven by both Hamiltonians repeatedly for different times: $\beta_i,\gamma_i$, which are the parameters of the circuit (see figure \ref{fig:QAOA}). Since the QAOA Hamiltonian is diagonal it can be decomposed as an Ising-like Hamiltonian whose evolution is easy to implement with the gate model as we explain in detail in the subsequent parts.
Also, diagonal Hamiltonians allow for the expectation value evaluation to be done efficiently on a classical device based on sampling the output of the circuit. Indeed, up to a chosen precision, this computation requires evaluating $h$ in given arguments corresponding to sampled circuit outputs,
\begin{equation}
\label{cost_func}
E(\vec \gamma, \vec \beta)=\braket{\psi(\vec \gamma, \vec \beta) | H_p | \psi(\vec \gamma, \vec \beta)} =\sum_z \abs{\braket{\psi(\vec \gamma, \vec \beta) | z}}^2 h(z) .
\end{equation}
The idea of Hardware Efficient Ansatz \cite{Kandala, Leone} expects that the complex enough Ansatz generates a state sufficiently similar to the ground state of the investigated Hamiltonian. In QAOA, by increasing the circuit depth -- number of parameterized gates -- the Ansatz allows us to achieve arbitrary closeness to the target state. Indeed, there exists a specific set of vectors $\Vec{\beta},\Vec{\gamma}$ such that the expectation value $E(\ket{\psi(\Vec{\beta},\Vec{\gamma})})$ approaches the target value when the depth increases. 
The desired set can be seen as a result of 'trotterization' of the adiabatic evolution from $H_0$ to $H_p$. Then the parameters $\Vec{\beta},\Vec{\gamma}$ should sum up to the value corresponding to the total running time. The approximation error for a fixed depth of QAOA was investigated for instance in \cite{akshay_circuit_2022,herrman_globally_2021,brandhofer_benchmarking_2022}.
\newline

Compressive sensing is a paradigm for recovering sparse signal $\bf x$ from small number of linear measurements represented as vectors ${\bf a}_i$ in incoherent basis. The measurement output are then given by $y_i={\bf a}_i^T\bf x$. Vectors ${\bf a}_i$ are called measurement patterns. Mathematically the recovery of $\bf x$ based on the information carried by pairs $\{{\bf a}_i,y_i\}$ is an under-determined problem, i.e., many vectors $\bf x$ fit to given measurements $\{{\bf a}_i,y_i\}$. Uniqueness of the solution can be specified demanding that the solution is the most sparse among all vectors that under the same set of measurements fit the data. It is indeed the correct solution if the number of measurements is sufficiently large with respect to the sparsity. Typically this number scales weakly with the dimensionality of the signal. So, if the linear measurement patterns ${\bf a}_i$ form rows of a transition matrix $A$, i.e., ${\bf y}=A\bf x$, where the sparse high dimensional signal is ${\bf x}$, and the outputs of the measurements form vector ${\bf y}$, 
the correct solution is given as 
\begin{equation}
    {\rm argmin}_{\bf x}\|{\bf x}\|_0 : \qquad {\bf y} = A{\bf x}.
\end{equation}
Here $\|\ .\ \|_0$ is the 0-norm. As this problem is NP-hard, \cite{muthukrishnan_data_2005} solutions are usually searched replacing the highest sparsity requirement, the minimum of 0-norm, by the minimum of $l_1$-norm. Many iterative greedy algorithms like for instance matching pursuit, orthogonal matching pursuit, etc. \cite{Mallatb, Zhangb, Blumensathb} are also developed to find the sparse solutions of the constraint problem. Under some conditions on the measurements and the sparsity, the alternative approaches can lead to correct or approximately correct solutions. 

There is a vast range of applications of compressive sensing \cite{romberg, mihajlovic, lustig, gross}. 
In \cite{roga_classical_2020,jacob_franck-condon_2020} it has been used as a tool in some sparse problems in the context of boson sampling and theoretical molecular spectroscopy. In \cite{roga_classical_2020} it has been shown how sparse distributions of boson sampling can be efficiently calculated by classical computers. The method can be applied for finding Franck-Condon coefficients \cite{gupta_2016, Doktorov,Jankowiak,guzik2015,huh,Nemeth,Oh} of vibronic spectra of large molecules. The proof of concept of this application was demonstrated in \cite{jacob_franck-condon_2020}. As discussed there, using the classical methods one can find spectra of molecules characterised by high sparsity, however the method stops working if the sparsity is too low. To deal with less sparse problems, we propose the extension of the classical method including a quantum subroutine for the crucial optimization in the algorithm which in general can be a problem of class NP. In this paper, for the first time, we demonstrate the proof of concept of this extension. 

\section{Quantum algorithm for compressive sensing}

Our goal is to recover a sparse or almost sparse distribution from a set of marginal distributions. In the typical case we are dealing with, the size of signal $\bf x$ is $d = 2^n$. So, entries of $\bf x$ can be indexed by $n$ bit strings. In our method we assume that the measurements ${\bf a}_i$ are binary patterns defined by fixed values of specific uplets of bits. It is a specific case of compressive sensing where we deal with a binary transition matrix and the measurements are given by structured easily to describe patterns. This allows us to overcome the problem of storing the measurement matrix $A$ which for a general random matrix requires a large amount of storage memory. The measurement characterised by patterns of fixed values of chosen bits indexing the entries of observed vector $\bf x$ appears naturally, for instance, in the case of an experiment with a set of Geiger mode detectors, where probabilities of all possible events form $\bf x$. The probability of a given event in a chosen set of detectors is equal to the sum of probabilities of all consistent events. Such marginal probability can be seen as summation of elements of $\bf x$ over patterns corresponding to fixed values of bits indexing the vector. Having such marginal distributions we can recover the full measured vector $\bf x$ given it is sufficiently sparse.

%

\begin{figure}[h]
    \centering
    \includegraphics[scale=0.28]{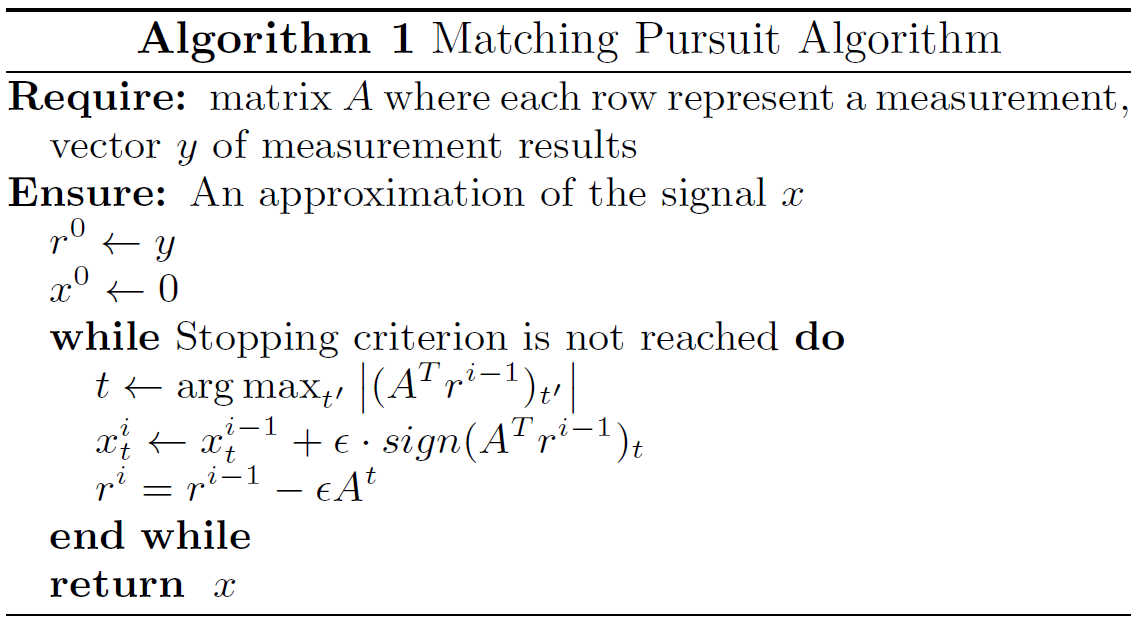}
    \label{algo1}
\end{figure}

We assume that we have access to the marginal distributions composing vector ${\bf y}$ which is the output from a set of measurements described by the measurement matrix $A$. The measurement matrix that transforms a joint distribution into marginal distributions consists of binary patterns ${\bf a}_i$ (rows of $A$) of size $d$. In order to recover an $s$-sparse signal, the number of patterns should be of the order of $s\log d$, however the specific number depends on the properties of the transition matrix and the recovery algorithm \cite{foucart_sparse_2013}. We are interested in using a basic algorithm called matching pursuit \cite{Mallatb} to recover $\bf x$ from the set of measurements. It is summarised in Algorithm 1. The computational bottleneck is the support detection stage in which a column of the transition matrix $A$ the most similar to the current residue of the marginal distribution needs to be determined. Formally, in the support detection stage of the $i$-th stage of the algorithm one needs to find 
\begin{equation}
\label{support_detection}
{\rm argmax}_t|A^T{\rm r}^{i-1}|_t.
\end{equation}

In the most general case, if we deal with a random measurement matrix of $m$ measurements and a sparse vector of size $N$: complexity of matrix-vector product is $\mathcal{O}(mN)$ while finding the largest element is $\mathcal{O}(N)$. However, in the case we are considering, the size of the signal scales as $N = \mathcal{O}(2^n)$ which exceeds the polynomial time complexity in both cases. This situation is intractable by classical computers and has a true interest in many domains as processing big data \cite{Cevher} or simulation of molecular spectra \cite{jacob_franck-condon_2020}.
In \cite{roga_classical_2020} we have shown that for specifically constructed measurements $A$, solving equation (\ref{support_detection}) is equivalent to finding the configurations of $n$ $\frac12$-spins that maximizes the energy of an interaction Hamiltonian. 
If we restrict the measurement patterns to the set with only neighbouring pairs of bits fixed, the Ising Hamiltonian that is obtained in this way can be solved by a classical algorithm in $poly(n)$, e.g., \cite{schuch_matrix_2010}. However, if we consider more complex patterns characterised by additional non-neighbour pairs, or larger uplets of bits fixed, the problem (\ref{support_detection}) becomes equivalent to the solution of the Ising model for spin-glass or generalized Ising model which is, in general, intractable efficiently by classical algorithms. As the quantum subroutine may extend the known class of solvable cases, we suggest to use it in the support detection stage of the matching pursuit protocol. The quantum solver allows us to extend the set of patterns of measurements beyond the fixed nearest neighbors bits patterns, increase the number of constraints, and in consequence, attack compressive sensing problems characterised by smaller sparsity. By using this hybrid setup we can still benefit from reduced storage and computational cost. 
\newline

The method consists in mapping the support detection optimization into an Ising Hamiltonian problem. To do so, we use the following framework. The signal vector $\bf x$ of size $d=2^n$ is sparse and hence contains only a few non-zero values which we call $(v_1,v_2,\ldots,v_s)$. Hence, the signal $\bf x$ can be written as $\textbf{x} = \sum_{i=1}^s v_i \textbf{e}_i$, where every standard basis element $\{\textbf{e}_i\}$ -- the spike position, can be uniquely characterised by a binary representation of $n$ bits, $\textbf{e}_i = b_1 b_2 \ldots b_n$. Moreover, we can represent every vector $\textbf{e}_i$ using the tensor product form: 
\begin{equation}
\textbf{e}_i = (1-b_1\quad b_1)_1 \otimes (1-b_2\quad b_2)_2 \otimes \ldots \otimes (1-b_n\quad b_n)_n.
\end{equation}
Next, our binary measurement patterns ${\bf a}_i$ can be written in the same notation. For a measurement characterised by the marginal distribution with bits $\{i_1,i_2,\ldots,i_k\}$ fixed as $\{b_{i_1},b_{i_2},\ldots,b_{i_k}\}$, the corresponding pattern ${\bf a}_i$ is:
\small
\begin{equation}
(1\quad 1)_1 \otimes \ldots \otimes (1-b_{i_1}\quad b_{i_1})_{i_1} \otimes \ldots \otimes (1-b_{i_k}\quad b_{i_k})_{i_k} \otimes \ldots \otimes (1\quad 1)_n.
\end{equation}
\normalsize

In this way the marginal distributions $\textbf{y}$ are now expressed as overlaps between measurement patterns $\textbf{a}_i$ and spike position vectors:
\begin{equation}
    \textbf{y}_i = \textbf{a}_i \textbf{x} = \sum_j v_j \textbf{a}_i \cdot \textbf{e}_j.
\end{equation}

This convention and the choice of the measurement matrix allow for two main advantages. First, we avoid the intractable problem of storing exponentially large unstructured measurement patterns of $A$. Second, products of $A$ with any sparse vector is now elementary. These two advantages will be useful latter to construct the Hamiltonian and use QAOA as the Ising solver.

Our quantum subroutine for support detection stage in Algorithm 1 uses the quantum device to solve a specific Hamiltonian. This Hamiltonian needs to represent our problem, i.e.,  (\ref{support_detection}). Also, due to the choice of the structured patterns this is the Ising Hamiltonian. Indeed, it reads, 
\begin{equation}
    \hat{H} = A^T y = \sum y_i \textbf{a}_i
\end{equation}
where $\textbf{a}_i$ are the rows of $A$ (columns of $A^T$). Every pattern $\textbf{a}_i$ can be represented in the conventional way in terms of Pauli matrices. Indeed, as they are tensor products of the terms $(0 \quad 1)$, $(1 \quad 0)$ and $(1 \quad 1)$, each of them (seen as a diagonal matrix) can be mapped to a combination of Pauli Z matrix and identity matrix as follows, 
\begin{align*}
    {\rm diag}((0 \quad 1))) &= \frac{I + Z}{2}\\
    {\rm diag}((1 \quad 0)) &= \frac{I - Z}{2}\\ 
    {\rm diag}((1 \quad 1)) &= I
\end{align*}
The mapping can be done efficiently in time $\mathcal{O}(nM)$, where $M$ is the number of measurement patterns.
Then, one can compute the tensor products to get a few Pauli strings that we add together weighted by the $y_i$.

The resulting Hamiltonian is a generalized Ising Hamiltonian, since it can contain terms that correspond to interactions between several spins depending on the number of fixed bits in the definition of our measurements, 

\begin{equation}
\label{hamiltonian}
    \hat{H} = \sum_{k}
    \sum_{\sigma_j \in \mathcal{P}^k} \alpha_j \sigma_j^Z,
\end{equation}
where $\mathcal{P}^k$ is the set of the Pauli strings of length $k$. The position of the maximum of such constructed Hamiltonian corresponds to (\ref{support_detection}), solving therefore the support detection problem in each step of Algorithm 1.

In general, Hamiltonians (\ref{hamiltonian}) are no longer tractable by classical methods contrary to the ones in 
\cite{roga_classical_2020, jacob_franck-condon_2020}. However, it is appropriate to be processed by a dedicated Ising solver, like an annealer or the QAOA algorithm. In the next section, we develop the method of finding the ground state of Hamiltonian $\hat H$ using QAOA.

\section{QAOA subroutine for Matching Pursuit}

As mentioned in the Introduction, to solve our Ising Hamiltonian -- and hence the support detection in Algorithm 1 -- we need to ensure two things to be done efficiently: implementing the Hamiltonian evolution on the quantum computer and evaluating the cost function.

In the most general case, the operations that one is able to implement on a quantum computer are composed of a set of universal gates. This set includes all possible single-qubit operations and at least one two-qubit gate. For example, IBMq qiskit API uses the set of $U(\theta,\phi,\lambda)$ gate \cite{noauthor_qiskit_nodate, cervera-lierta_exact_2018}, that can reproduce every single-qubit rotation in the Bloch sphere, and the two-qubit CNOT gate. To implement the unitary evolution with the problem Hamiltonian (\ref{hamiltonian}), we need to be able to represent it as a series of quantum gates within such a set. Explicitly, this unitary transformation
\begin{equation}
    U(t) = e^{i t \hat{H}} = e^{i t \sum_k \sum_{\sigma_j \in \mathcal{P}^k} \alpha_{j} \sigma_j},
\end{equation}
using 'trotterization' technique, can be written as a product of short evolutions of the form: $e^{i t \alpha_j \sigma_j}$ where here, $\sigma_j$ can denote a product of a few Pauli-Z operators applied to different qubits. For evolution driven by a single, $k=1$, Pauli operator, the implementation is feasible in terms of a single qubit $Z$-rotation gate: $R_Z(t\alpha) = e^{i t \alpha \sigma}$. Similarly, for the evolution driven by $k=2$ local Pauli-Z operators, we use a two-qubit $ZZ$-interaction $R_{ZZ}(t\alpha) = e^{i t \alpha \sigma_{j_1} \sigma_{j_2}}$. This interaction is not in the quiskit universal set but can be rewritten as two CNOT gates and one $R_Z$ single-qubit rotation: $R_{ZZ} = CX \cdot R_Z \cdot CX$.

By generalizing this procedure, any evolution including a string of $k$ Pauli-Z operators can be written in the gate model using $2(k-1)$ CNOT gates and one single-qubit rotation $R_Z$. We use this method to implement our problem Hamiltonian (\ref{hamiltonian}) in the QAOA algorithm. 
As the mixing Hamiltonian we can chose the sum of local Pauli-$X$ operators the evolution of which is also implementable with $R_X$-rotations gates.

The second essential point to consider in the QAOA algorithm is the evaluation of the cost function $h$ that should be done efficiently on a classical computer. After running the Anzats quantum circuit, we sample from the output state $\ket{\psi(\vec \gamma, \vec \beta)}$ and for every element $\bra{z}$ representing a location of the non-zero signal in $\bf x$ in binary notation
, we evaluate the cost according to equation (\ref{cost_func}). The function $h(z) = \abs{(A^T)_z y}$ needs to be efficiently calculable for any chosen $z$. This is actually the case. Indeed, according to the choice of our matrix $A$, $(A^T)_z$ is not stored but can be computed efficiently and so is the product with $y$ (see \cite{roga_classical_2020} for details).
\newline





In summary, the whole process goes as follows:
\begin{itemize}
    \item Run Algorithm 1 on classical device until support detection part. Then for every support detection step:
    \item Formulate the Hamiltonian of equation (\ref{hamiltonian}) following the previous procedure.
    \item Decompose its time-evolution operator into rotation gates to build a circuit.
    \item Build the QAOA circuit, repeating the layers with a number of free parameters (that is to determine).
    \item Optimize the circuit with classical methods and obtain the solution (the eigenvector corresponding to the maximum eigenvalue).
    \item Continue running Algorithm 1, repeating all steps with new residues until a stopping criterion is met.
\end{itemize}




\section{QAOA for compressive sensing -- simmulation results}


In what follows, we explain our results based on simulations. The goal is to investigate whether our method described above offers any advantage over the classical method that has been used before, mainly in \cite{roga_classical_2020} and \cite{jacob_franck-condon_2020}. The advantage of the method elaborated here is supposed to come from the more constrained Hamiltonian (multi-spin, non-local spins interactions) than the Ising 
spin-chain Hamiltonian formed with nearest neighbour interaction patterns used in \cite{roga_classical_2020,jacob_franck-condon_2020}. This implies that one can use more measurements with more varied measurement patterns without strong repercussions on computation time due to the quantum solvers. Indeed, additional measurements and freely chosen measurement patterns should lead to better determined constraints appropriate to solve problems with smaller sparsity.

In what follows, the plots are done based on our implementation of QAOA algorithm using qiskit library \cite{noauthor_qiskit_nodate-1}. There are two motivations for choosing QAOA rather than other methods as, e.g., Annealing in this research: recent research \cite{zhou_quantum_2020} indicates that QAOA should have a better generalization power and be able to learn through non-adiabatic processes if well initialized, allowing to solve even small-gap Hamiltonians; QAOA simulates the evolution of the Ising Hamiltonian making it, in theory, possible to implement any Hamiltonian while some Hamiltonian evolutions may not be feasible in the annealing scenario. Last but not least, the number of qubits needed by our algorithm scales well, like the logarithm of the problem size.

\begin{figure}[!h]
\subfloat[Original spectrum]{\includegraphics[scale=0.5]{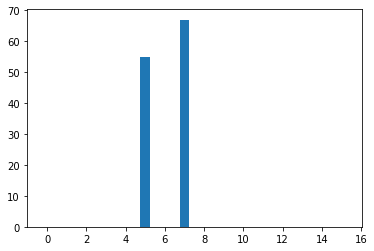}}
\subfloat[Classical reconstruction -- brute force optimizer]{\includegraphics[scale=0.5]{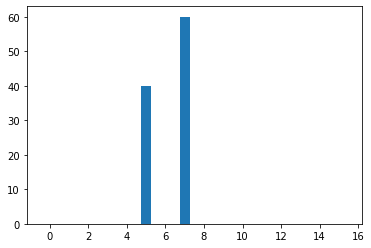}}\\
\subfloat[nearest neighbor constraints -- classical solver]{\includegraphics[scale=0.5]{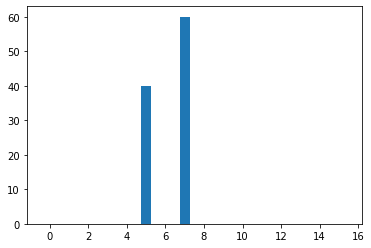}}
\subfloat[nearest neighbor constraints -- QAOA solver]{\includegraphics[scale=0.5]{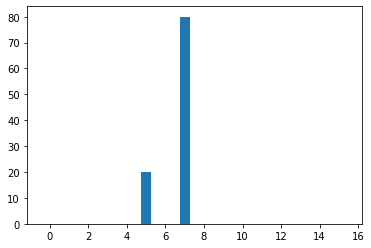}}

\caption{Example of performance of three optimization methods in Algorithm 1 aimed at recovering the randomly chosen sparsity 2 spectrum over $6$ bits space shown in part (a). In part (b), the spectrum reconstructed with brute force optimizer. In part (c), reconstruction with classical solver \cite{roga_classical_2020} based on constraint optimization equivalent to the solution of a nearest neighbor Hamiltonian Ising problem. In part (d), reconstruction based on optimization QAOA solver of nearest neighbor Ising Hamiltonian. In this first case, QAOA with nearest neighbor patterns performs as well as both classical algorithms being able to recover the whole spectrum.}
\label{fig:perfom_as_well}
\end{figure}

\begin{figure}[!h]
\subfloat[Original spectrum]{\includegraphics[scale=0.5]{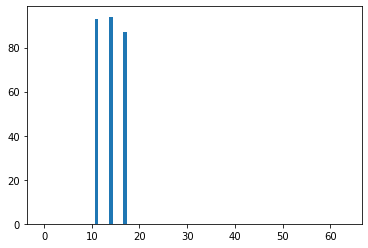}}
\subfloat[Classical reconstruction -- brute force optimizer]{\includegraphics[scale=0.5]{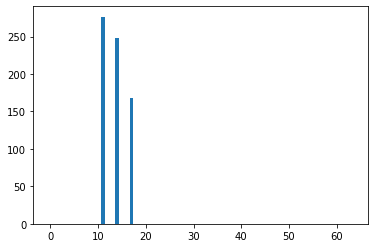}}\\
\subfloat[nearest neighbor spins measurements -- classical solver]{\includegraphics[scale=0.5]{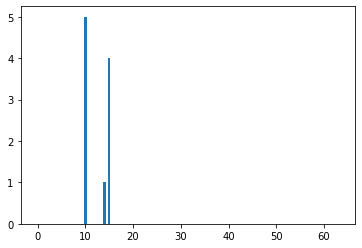}}
\subfloat[Constraints involving quadruplets of spins measurements -- QAOA reconstruction]{\includegraphics[scale=0.5]{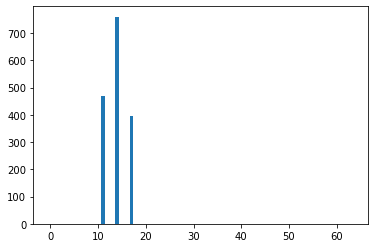}}

\caption{Example of performance of three optimization methods in Algorithm 1 aimed at recovering the randomly chosen sparsity 3 spectrum over $6$ bits space shown in part (a). In part (b), the spectrum reconstructed with brute force optimizer. In part (c), reconstruction with classical solver \cite{roga_classical_2020} based on constraint optimization equivalent to the solution of a nearest neighbor Hamiltonian Ising problem. In part (d), the spectrum reconstructed by QAOA solvers of Hamiltonians corresponding to interaction between randomly chosen quadruplets of spins. In this example the quantum method outperforms the classical one based on exact solutions of nearest neighbor interaction Hamiltonian. Indeed, the classical method was unable to recover the whole spectrum while the QAOA one did it well.}
\label{fig:outperform}
\end{figure}

Here, we compare our quantum solver-based implementation with the classically efficient method of \cite{roga_classical_2020} which is based on the nearest neighbour spin interaction Hamiltonian. We show results in figures \ref{fig:perfom_as_well} and \ref{fig:outperform}. 
To test both approaches, classical and quantum, we generate a random spectrum of a given sparsity choosing the position of each pick from a uniform distribution. An example is shown in figure \ref{fig:perfom_as_well} (a). Next we apply appropriate measurement patterns and recover the spectra according to Algorithm 1 with classical or quantum solver in the support detection stage (\ref{support_detection}). The results of the reconstruction are shown in parts (c) and (d) of figure \ref{fig:perfom_as_well} respectively. In both cases the same nearest neighbor interaction Hamiltonians were solved. For a comparison, in figure \ref{fig:perfom_as_well} (b), we also show the performance of the algorithm with a conventional brute force optimization method in the support detection stage which, in general, takes exponential time. In our comparison of different methods, we focus on the ability of an algorithm to correcly find positions of all spikes in the spectrum. Here, the value of each spike is more related to classical settings of the matching pursuit algorithm which influence all optimization methods, so the problem of the values of the spikes is secondary with respect to the problem of localizing their positions.

\onecolumngrid

\twocolumngrid

In figure \ref{fig:perfom_as_well}, all spikes have been recovered in both cases, i.e., with both classical and quantum method. This shows that there exist cases where our quantum algorithm is able to perform as well as the classical algorithm based on the nearest neighbor Ising solver studied in \cite{roga_classical_2020}. The observation from figure \ref{fig:perfom_as_well} does not imply that the quantum algorithm will perform as well every time. However, the observation is already optimistic for QAOA-based algorithm. Indeed, if we run the quantum algorithm with measurements corresponding to the nearest neighbor interaction patterns, we can perform at most as well as the classical method, since, while the amount of information about the spectrum brought by the measurements is the same in both cases, the classical method being the exact method makes an optimal use of this information. QAOA based method being an approximate method with the constraints used here is in a worse starting position. 

Figure \ref{fig:outperform} shows a different case. Here our quantum algorithm has been able to reconstruct the original spectrum more accurately than the classical algorithm of \cite{roga_classical_2020}. This shows that there exist cases where our quantum algorithm may outperform the classical method.
We assume it is only possible since we provide the quantum algorithm with more complex than the nearest neighbour interaction patterns constraints, bringing more information about the spectrum. Even though one can perceive here an advantage for the quantum algorithm over classical methods, it does not mean that using more complex measurements will allow the quantum algorithm to reconstruct any spectrum or even to outperform the exact classical nearest neighbor constraints based method every time. The quality of the reconstruction depends on two things: the information contained in the initial measurements and the ability of our algorithm to exploit this information. The latter, for our QAOA algorithm, reflects the choice of hyper-parameters and settings, e.g., the number of free-parameters, the optimizer, etc.

\newpage
\onecolumngrid


\begin{figure}[!h]
\subfloat[Performance for nearest neighbor Measurements]{\includegraphics[scale=0.5]{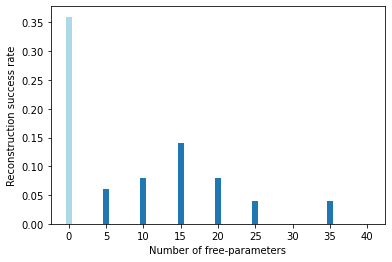}}
\subfloat[Performance for Quadruplets Measurements picked randomly]{\includegraphics[scale=0.5]{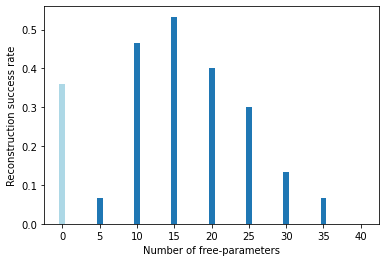}}

\caption{Rate of successful reconstruction by QAOA scheme for spectrum drawn randomly for different number of free-parameters. Light-blue spike on the right of each plot is the success rate for classical nearest neighbor method. (a) shows reconstruction success rate on several random spectrum using nearest neighbor measurements solved by QAOA scheme for different parameters. (b) shows reconstruction success rate on several random spectrum using randomly choosen quadruplets measurements solved by QAOA scheme for different parameters.}
\label{fig:success_rate}
\end{figure}

\twocolumngrid

In what follows, we analyze the reconstruction success rate of the quantum algorithm compared with the classical one. First, we study the case when we feed the algorithm with the nearest neighbor-type measurement outputs, then the case when measurements corresponding to randomly chosen quadruplets are used. For the reasons already mentioned, we expect that in the first scenario QAOA will never outperform classical nearest neighbor Ising Hamiltonian solver, implying that the success rate should remain not better than in the case of the classical solver. In the second case, QAOA may sometimes outperform the classical algorithm implying the higher or lower success rate depending on the randomly chosen measurement constraints or the choice of hyper-parameters. So, we ask if providing more complex measurements allows us to observe cases where QAOA outperforms the classical method based on exact solution of the nearest neighbor Ising Hamiltonian. In figure \ref{fig:success_rate} we show the results of our analysis regarding the reconstruction success rate.

The analysis was performed as follows. We randomly selected 
random spectra of sparsity 3 over 6 bits space. We simulated the appropriate measurements and fed the Algorithm 1 with the measurement results trying to reconstruct the original spectra when the classical or quantum algorithm was used as the subroutine of the algorithm. First we used nearest neighbor-type measurements in both, classical and quantum case. 
Then, we reconstructed the spectra using the QAOA-based algorithm with randomly selected measurements of quadruplets of fixed bits which reduced our problem to the solution of the Ising Hamiltonian with four, not necessary neighbouring spins interactions. We repeated the quantum algorithm with different numbers of free-parameters. Finally, in each case, we counted the number of successful reconstructions, i.e., the number of times the positions of all lines of the spectrum have been correctly recovered.

We show the results in figure \ref{fig:success_rate}. On the left of each plot in light-blue, we show the success rate of the classical algorithm based on the nearest neighbor-type measurements which is found to be around $36\%$ for randomly chosen spectra. For the QAOA with the nearest neighbor-type measurements, \ref{fig:success_rate} (a), the reconstruction success rate is always lower than using the classical method with a maximum at around $14\%$ which attests our first observation: as we use the same measurements, the amount of information about the spectrum brought to the algorithms is the same and so, the quantum algorithm can only perform as well or worse than the classical one. 

However, when we feed the QAOA with the information based on quadruplet-type measurements it may have access to greater amount of information about the investigated spectrum than with only nearest neighbor-type measurements. We observe that, depending on the setting, it is possible to outperform the classical method with the nearest neighbor constraints on the success rate of randomly selected problems. Indeed, in figure \ref{fig:success_rate} (b), one can see that using a right number of free-parameters (the best seems to be around $15$) the quantum algorithm based on quadruplets-type measurements, achieving around $53\%$ success rate, is able to reconstruct the spectrum in more cases than the classical method. 

\section{Conclusion}

In this work we investigated the feasibility of using quantum computers for compressive sensing problems. We recognized that in the reconstruction of highly dimensional sparse vectors from logarithmic number of marginal measurements quantum computers playing the role of optimizers can offer an advantage. In particular, we tested the type of measurements for which the optimization in the matching pursuit subroutine which is the bottle neck of the algorithm reduces to the solution of an Ising problem. This, in general, belongs to the class of NP-complete problems. However in particular cases, restricted to a selection of measurements in the compressive sensing approach, this optimization can be efficiently solved by classical methods (solving nearest neighbor interaction Hamiltonians) as it was shown in \cite{roga_classical_2020, jacob_franck-condon_2020} in the context of reconstruction of sparse output configurations of large interferometers. The sparsity of classically tractable problem was however required to be high which limited applicability of this approach. 

Adding more measurements allowed us to relax the high-sparsity restriction and extend the class of tractable problems. However this implied adding non-local and many-body interaction terms to the Hamiltonian whose ground state needed to be found. With this step we entered to the domain of spin glass problems where the ground state is not solvable classically. It might be not solvable by quantum computers neither, as it is strongly believed that NP complexity class is not included into the BQP class. However adding new selected constraints gradually and in a controllable way may lead to the area where quantum computers acting on different principles than classical algorithms may still operate, while the known classical approaches fail. Typically non NP-complete problem from the NP class that would be inside the BQP class.     

In this study we simulated the quantum approximate optimization algorithm (QAOA) as a solver of a constraint optimization problem. We shown that while this algorithm cannot guarantee the exact solutions, it loses to the classical solvers in the domain where the latter operates. However, the quantum algorithm still provides meaningful solutions in the regime where the classical ones show their insufficiency. Indeed, even with merely randomly chosen measurement patterns and sub-optimal optimization of QAOA, we were already able to solve a large number of new randomly selected spectra with an around $20\%$ increase in the success rate with respect to the classical Ising solver based on nearest neighbor-type measurements.

This work was motivated by the previous research devoted to vibrational spectra of molecules \cite{gupta_2016,guzik2015,jacob_franck-condon_2020}. Meanwhile this problem was satisfactorily solved in \cite{Oh}, in the present research, we signalize the applicability of the studied method in much wider class of compressive sensing challenges that may include big data tasks like, for instance, PageRank or quantum tomography. In the latter case we expect the relevant Hamiltonians not built of Pauli-Z operators only. Indeed, recognizing and characterising classes of large sparse signals which may benefit the most from what quantum solvers offer is left as the subject for the future work.\\

\textbf{\emph{Acknowledgements}}
This work was supported by JST COI-NEXT Grant No. JPMJPF2221. W.R. and M.T. were also supported by JST Moonshot R\&D Grant No.~JPMJMS2061 and No.~JPMJCR1772. W.R. thanks The Erwin Schrödinger International Institute for Mathematics and Physics (ESI) which hosted him during the Thematic Program: Quantum Simulation - from Theory to Application, Sep. 2019, where some ideas of this paper were developed. 

\bibliography{ref}

\end{document}